\documentclass[11pt]{article}

\usepackage{array}
\usepackage[a4paper,margin=2.5cm]{geometry}
\usepackage[T1]{fontenc}
\usepackage{lmodern}
\usepackage{microtype}
\usepackage{amsmath,amssymb}
\usepackage{graphicx}
\usepackage{booktabs}
\usepackage{subcaption}
\usepackage{xcolor}
\usepackage[numbers,sort&compress]{natbib}
\usepackage[hidelinks]{hyperref}
\usepackage[capitalise,noabbrev]{cleveref}

\title{One Neural Operator for Re-planning Low-Thrust Cislunar Transfers after Thrust Outages,\\
at the Original or a Later Arrival}
\author{Selim Olgu Pilav\\ \small Independent researcher\\ \small \texttt{sop24@cantab.ac.uk}}
\date{}

\begin{document}
\maketitle

\begin{abstract}
Safe-mode events causing thrust outages are common in missions with electric propulsion. To demonstrate mission feasibility in such an event, missed-thrust margin studies are performed where thousands of outage scenarios need an exact re-plan solved. We train a neural operator that maps a nominal trajectory with an outage profile to a re-planned trajectory and to the probability that the recovery exists, for 270 low-thrust near-rectilinear halo orbit (NRHO) to distant retrograde orbit (DRO) transfers, at an allowed slip of up to 10 days. On unseen transfers and zero slip, its median control error on the severe half of outages is 1.20\% compared to 14.0\% for the best cheap re-planner benchmark. With the arrival slip as an input, the error is 1.67\% against 23.2\%. It also predicts the recoverability of an outage for a given arrival slip with an area under the receiver operating characteristic curve (AUC) of 0.997, while the minimum recoverable slip prediction is optimistic by more than half a day on only 6 of the 84 checked outages needing slip. At the original arrival, the ensemble spread of an operator variant with solver inputs has a rank correlation of 0.91 with the actual error, and performing exact solves on the 31\% most uncertain severe outages reduces the 90th-percentile error from 8.5\% to 2.3\%. A query takes milliseconds without a solver, making this a useful tool for large outage studies where the cost of thousands of exact solves outweighs the time to train the operator.
\end{abstract}

\section{Introduction}\label{sec:intro}
Safe-mode events are common in deep-space missions, including missions that use electric propulsion. These events create outages in thrust, and then the transfer must recover with the remaining time and fuel. Flight records of about 240 safe-mode events give Weibull distribution fits for the duration and spacing of safe-mode events \citep{imken2018}. In April 2025, a single anomaly cost Psyche 76 days of thrusting \citep{nasa2025psycheA,nasa2025psycheB}. At the design stage of low-thrust missions, $\Delta v$ and arrival slip margins are set by sweeping outages along the nominal trajectory and re-optimising a recovery for each case \citep{laipert2015}, or by a Monte Carlo over flight statistics \citep{laipert2018}. For example, the Dawn mission trajectory and margins were designed to tolerate a 28-day shutdown at any point \citep{rayman2006}. Psyche, which uses Hall-effect thrusters in deep space, set explicit missed-thrust requirements stating that a 10-day outage at any time and a 22-day outage during solar conjunctions must be tolerable \citep{oh2019psyche}.

In these large outage studies, every sampled outage is a full re-optimisation. The Gateway insertion analysis re-optimised a recovery from three missed-thrust cases on each of 346 reference trajectories, over 1{,}000 solves, each with SNOPT inside a basin-hopping search \citep{karn2024}. Cislunar space is gaining traction with missions such as CAPSTONE, which entered a near-rectilinear halo orbit (NRHO) in November 2022 \citep{agasid2024capstone}, Artemis~I, whose Orion spacecraft flew a distant retrograde orbit (DRO) \citep{williams2023artemis}, and the Gateway, whose reference orbit was the 9:2 L2 southern NRHO \citep{lee2019gateway}, planned to be reached by an electric propulsion system \citep{karn2024} before NASA paused it in March 2026 \citep{vigliarolo2026}. Transfers between an NRHO and a DRO \citep{zimovan2017} are also a current research topic \citep{mccarty2020,karn2021,zimovanspreen2022}. For the NRHO-to-DRO transfers in this study, the exact labelling procedure takes 14.5~s at the median and 71~s on average. A rapid tool that outputs a re-planned trajectory with accurate $\Delta v$ margins and closest lunar approach, together with an indication of whether a recovery exists at a given arrival slip, would accelerate outage studies.

Several methods build robustness to missed thrust into the nominal trajectory \citep{mccarty2020,venigalla2022}. In \citep{mccarty2020}, a few branches of trajectories are optimised together with an NRHO-to-DRO reference, cutting the extra propellant needed after a 7-day outage by up to 90\%. The authors of \citep{venigalla2022} apply a similar idea to interplanetary transfers, where they optimise the nominal trajectory together with a group of recovery trajectories. However, these studies add a set of chosen outages to one optimisation problem, so a new nominal trajectory would need a new solve. The authors of \citep{sinha2025} bound the missed-thrust duration for which a first-order model of the recovery stays valid along the nominal trajectory. This results in a certificate that shows when the recovery problem is well behaved but does not fully determine if a recovery exists.

Learned methods compute optimal solutions in advance and train a network on them, so that each re-plan takes only one network evaluation. In \citep{rubinsztejn2020}, the authors train a network on optimal state--control pairs to fly as a feedback law, and the network repairs more than 75\% of missed-thrust events. Similar networks can give real-time guidance for landing and for low-thrust transfers \citep{sanchez2018,izzo2021}. However, these policies are generally compared to the exact solutions, not against cheap re-planners which also start from the nominal trajectory. Therefore, a network could appear accurate even when a cheap solver step from the nominal would do as well. In addition, these networks do not say whether a recovery exists, which is the critical information a margin study needs prior to using a re-planned trajectory. Thus, it is not clear if a learned re-planner performs better than a single convex or sequential quadratic programming (SQP) step from the nominal trajectory. Neural operators learn maps between functions \citep{li2021fno,lu2021deeponet}, which matches the form of this problem where the input is the thrust available over time and the nominal trajectory, and the output is a re-planned thrust history from the moment thrust recovers.

In this work, we train one neural operator for a transfer family consisting of 270 low-thrust NRHO-to-DRO transfers. The neural operator returns the re-planned trajectory and the probability that a recovery exists. The operator is an ensemble of Fourier neural operators \citep{li2021fno} that are trained on exact trajectory re-plans. It is queried with the thrust available over time and the nominal trajectory. The controls flown until thrust returns are frozen, and later outages are unknown. Therefore, the re-plan is causal and targets the trajectory post thrust recovery. We compare the operator against cheap re-planners on transfers unseen in training, with the main comparisons fixed in advance.

The paper investigates three questions:
\begin{enumerate}
  \item Fixed arrival: The operator re-plans each outage with no allowed arrival slip, and its control error on the more severe half of recoverable outages is compared with that of the best cheap re-planner.
  \item Recoverability: The operator predicts whether a recovery on time exists. The same output is used to derive the longest outage a transfer can survive at each point along the trajectory.
  \item Late arrival: A single operator takes the arrival slip as an input and re-plans for arrival slips up to 10 days late. Its minimum recoverable slip is compared against exact searches on 200 outages.
\end{enumerate}

\section{Problem}\label{sec:problem}

\begin{figure}[!t]
  \centering
  \begin{subfigure}[b]{\textwidth}\centering
    \includegraphics[width=0.92\textwidth]{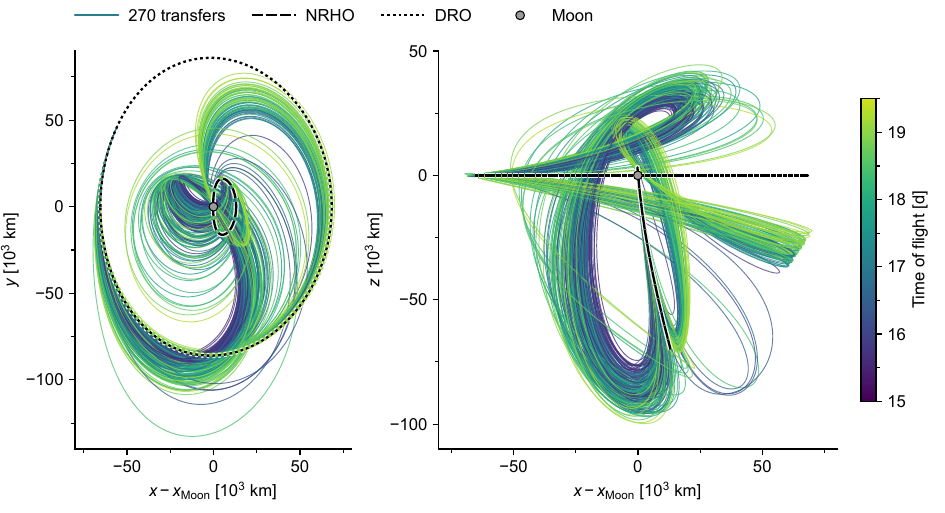}\caption{}
  \end{subfigure}\\[4pt]
  \begin{subfigure}[b]{0.55\textwidth}\centering
    \includegraphics[width=\textwidth]{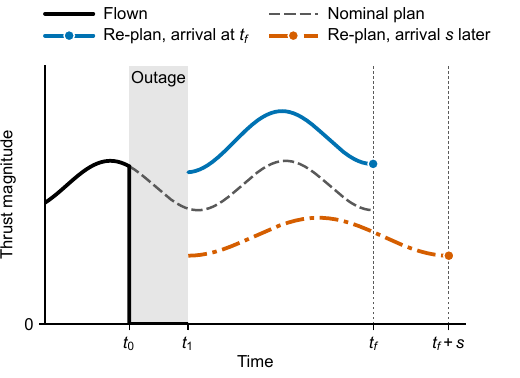}\caption{}
  \end{subfigure}
  \caption{(a) The 270 NRHO-to-DRO low-thrust transfers in the Earth--Moon rotating frame. (b) Thrust along one transfer with an outage from $t_0$ to $t_1$.}
  \label{fig:problem}
\end{figure}

We consider NRHO-to-DRO transfers. Each transfer leaves the 9:2 L2 southern NRHO and arrives on a DRO whose Earth-side crossing lies 0.17~LU from the Moon, about 65{,}000~km, where LU is the Earth--Moon distance. Both orbits are calculated in the Earth--Moon circular restricted three-body problem (CR3BP). The transfers are then flown in the bicircular restricted four-body problem (BCR4BP), adding the Sun. We sample 270 minimum control-energy transfers with a Sobol sequence over the transfer time, the departure and arrival phases and the Sun phase (\cref{fig:problem}a). The Sobol sequence ensures even sampling \citep{sobol1967,owen1998}. The family's transfer times span 15.4--19.3~d, but the test transfers are drawn from an envelope of 15.6--19.1~d, ensuring every test transfer lies within the training range. The problem is made tractable by setting a constant mass and thrust acceleration bound of 0.82~mm/s$^2$. The optimisation minimises the control energy for each transfer time at a fixed time of arrival while keeping 100~km above the lunar surface. Each problem is transcribed by multiple shooting on 200 nodes with a terminal error below 1~km (Appendix~\ref{app:problem}).

The outage structure is derived from the safe-mode records of \citep{imken2018}, which have a Weibull fit for the duration of a safe-mode event with a shape of 1.17 and a scale of 2.41~d. Every outage is a full shutdown, completely stopping the thrust. The outage duration is drawn with the same shape but with the scale multiplied by 1.5 and kept within 0.5--10~d. Additionally, a second and a third outage occur with probabilities of 0.15 and 0.03. To vary the thrust history beyond full outages, the available thrust also falls linearly over the transfer in 30\% of the histories, from full thrust at departure to a loss drawn uniformly up to 20\% at the original arrival time. This modelling choice is added to replicate a gradual decline in the power or thruster performance and to create greater variability in thrust histories. Unlike the second and third outages, it is known at re-plan time. Due to the longer durations, the training outages are harsher than flight, so the hard cases are well sampled. Each case also stores the density ratio between the flight and training distributions, allowing metrics to be re-weighted to flight-like frequencies of outages lasting 0.5--10~d, with the thrust decline left as drawn (Appendix~\ref{app:problem}). It is also worth mentioning that in flight the probability of an outage occurring during a 17-d transfer is about 7\% \citep{imken2018}, and every result here is conditional on one occurring.

The re-plan starts where thrust returns and must still reach the DRO, either at the original arrival time $t_f$ or $s$ days later, where $s$ is the allowed arrival slip. Until the end of the first outage, the spacecraft flies the nominal trajectory with the controls clipped to the available thrust and coasts during the outage. From the point where thrust returns, the re-plan minimises the control energy to the DRO arrival state at the original final time $t_f$ or the specified slip, within the available thrust and with later outages unknown. An outage is deemed recoverable if the exact solver (\cref{sec:rivals}) finds a re-plan. We impose no propellant limit, so the recoverability depends only on the remaining time and thrust. We run two studies called the fixed-arrival study, where no slip is allowed, and the late-arrival study, where $s$ is drawn uniformly in 0--10~d. For the no-arrival-slip case, the outages start within 5--50\% of the transfer, as later outages are mostly unrecoverable without slip. For the late-arrival study, the outage occurrence is extended to start as late as 75\% of the transfer, and many of the outages that are unrecoverable on time can become recoverable. In both studies, the operator returns the re-planned states and controls, as well as the probability that recovery exists. The two studies are compared in \cref{tab:setup}.

\begin{table}[t]
\centering
\caption{The two studies.}
\label{tab:setup}
\small
\renewcommand{\arraystretch}{1.15}
\begin{tabular}{@{}>{\raggedright\arraybackslash}p{0.19\textwidth}
                >{\raggedright\arraybackslash}p{0.375\textwidth}
                >{\raggedright\arraybackslash}p{0.375\textwidth}@{}}
\toprule
\multicolumn{3}{@{}l}{\emph{Shared}} \\
Dynamics & \multicolumn{2}{>{\raggedright\arraybackslash}p{\dimexpr0.75\textwidth+2\tabcolsep\relax}@{}}{Bicircular restricted four-body problem} \\
Transfers & \multicolumn{2}{>{\raggedright\arraybackslash}p{\dimexpr0.75\textwidth+2\tabcolsep\relax}@{}}{270 minimum control-energy NRHO$\to$DRO transfers, time of flight 15.4--19.3~d, constant thrust acceleration bound of 0.82~mm/s$^2$} \\
Transcription & \multicolumn{2}{>{\raggedright\arraybackslash}p{\dimexpr0.75\textwidth+2\tabcolsep\relax}@{}}{200 nodes, terminal error below 1~km, at least 100~km above the lunar surface} \\
Split & \multicolumn{2}{>{\raggedright\arraybackslash}p{\dimexpr0.75\textwidth+2\tabcolsep\relax}@{}}{210 training, 30 validation and 30 test transfers, validation and test transfers inside a 15.6--19.1~d envelope} \\
Outages & \multicolumn{2}{>{\raggedright\arraybackslash}p{\dimexpr0.75\textwidth+2\tabcolsep\relax}@{}}{Full shutdowns of 0.5--10~d, 1.5 times longer than in flight \citep{imken2018}, up to three per transfer, a thrust decline of up to 20\% in 30\% of thrust histories} \\
\midrule
 & \textbf{Fixed arrival} & \textbf{Late arrival} \\
Arrival & Original time and DRO state & Slip $s$ drawn uniformly in 0--10~d \\
Outage start & 5--50\% of the transfer & 5--75\% of the transfer \\
Exact re-plans & 6{,}750 & 13{,}500 \\
Operator & Mixed operator: ensemble of 10 correcting R1 and R3 & Joint operator: ensemble of 10 correcting the nominal\\
Rivals & R0--R3 and R5--R7 & R0--R2 and R5--R6 \\
Severe outages & Recoverable outages removing at least 51~m/s of planned median $\Delta v$  & Recoverable outages removing at least 61~m/s of planned median $\Delta v$ \\
\bottomrule
\end{tabular}
\end{table}

\section{Method}\label{sec:method}

\subsection{Rivals}\label{sec:rivals}
We compare the operator to cheap rivals that start from the same nominal trajectory and take up to two solver steps. R0 flies the nominal trajectory with the controls clipped to the available thrust, with no re-plan. R1 takes one convex step about the nominal, and R2 takes two successive convexification (SCvx) iterations \citep{malyuta2022}. R3 takes one exact SQP step from the nominal trajectory and its multipliers. R3 is the first-order approximation of the optimum re-plan after an outage, so it tends to be accurate after mild outages. R3 is not valid for the late-arrival study as the extended trajectory has no multipliers (Appendix~\ref{app:rivals}). R5--R7 are devised to give the rivals the same training data as the operator, and they add a learned correction to R1--R3, respectively. The corrections to R1--R3 are predicted by extra trees \citep{geurts2006} from the outage description, fitted on the training and validation transfers. The outage description holds the start and length of each outage, the thrust decline and the transfer's time of flight and phases. The rivals are numbered so that R($k$+4) is R$k$ with the applied learned corrector. R0 has no learned correction, as it does not re-plan, meaning no R4 exists.

The exact re-plans, referred to as labels, are generated with IPOPT \citep{wachter2006ipopt}. These are started in two ways: one from the nominal trajectory with its Lagrange multipliers and one from the rival R1. The solution is deemed to be unique when both starts converge, and the relative $L_2$ norm of the controls agrees to within 0.1\%. The solution with lower cost is kept.

\subsection{Operator}\label{sec:operator}

We use a Fourier neural operator \citep{li2021fno} on the 200-node grid, which maps input histories to output histories. At each node, the input holds the available thrust, a flag marking the nodes already flown, the nominal controls $\mathbf{u}^{\mathrm{nom}} \in \mathbb{R}^3$ and states $\mathbf{x}^{\mathrm{nom}} \in \mathbb{R}^6$, the time as a fraction of the total transfer time, the time of flight and the Sun phase. The controls of R1 and R3 are also received at each node by the members of the mixed operator that use them. The inputs at each node are first mapped to 64 internal features. These then pass through four Fourier layers,
\begin{equation}
  \mathbf{h}^{(\ell+1)} = \sigma\!\left(W_\ell\,\mathbf{h}^{(\ell)}
  + \mathcal{F}^{-1}\!\left[R_\ell \cdot \mathcal{F}\mathbf{h}^{(\ell)}\right]\right), \qquad \ell = 0,\dots,3,
  \label{eq:fno}
\end{equation}
where $\mathcal{F}$ is the discrete Fourier transform over time, $W_\ell$ is a pointwise linear map and $\sigma$ is the Gaussian error linear unit (GELU) activation. $R_\ell$ multiplies the 32 lowest-frequency modes by learned weights and discards the rest. The output of the last layer is projected pointwise to a correction of a base trajectory, $\hat{\mathbf{u}} = \mathbf{u}^{\mathrm{b}} + \Delta\hat{\mathbf{u}}$ and $\hat{\mathbf{x}} = \mathbf{x}^{\mathrm{b}} + \Delta\hat{\mathbf{x}}$. The states are predicted directly by the operator rather than integrated from the control. This is done because, in the sensitive Earth--Moon--Sun dynamics, small errors in the controls grow quickly when integrated.

During training, each member minimises
\begin{equation}
  \mathcal{L} = \frac{\lVert \hat{\mathbf{u}} - \mathbf{u} \rVert_{\mathcal{R}}}{\lVert \mathbf{u} \rVert_{\mathcal{R}}}
  + \lambda_x \operatorname{MSE}\!\left(\Delta\hat{\mathbf{x}},\, \mathbf{x} - \mathbf{x}^{\mathrm{b}}\right),
  \label{eq:loss}
\end{equation}
where $\mathbf{u}$ and $\mathbf{x}$ are the controls and states of the label, $\lVert\cdot\rVert_{\mathcal{R}}$ is the $L_2$ norm over the set $\mathcal{R}$ of re-planned nodes, MSE is the mean squared error, the states are normalised per node and $\lambda_x = 0.1$. In the control term, the nodes prior to the outage end are not included.

The base trajectory can either be the nominal trajectory itself, or a rival's re-planned trajectory. This choice, together with what the operator outputs, results in the three operators in this paper. First, the mixed operator has ten members, five correcting R1 and five correcting R3. This operator gives the main fixed-arrival results. For this operator, both rivals must be computed before every query, and it only exists for the fixed-arrival study. On the other hand, the no-solver operator corrects the nominal directly, so a query needs no solver. This exists only for the fixed-arrival study. Finally, the joint operator adds one more output to the no-solver operator, the probability that a recovery exists, and it is the only operator type used in both studies.

For this, the network has an extra output, whose mean over the nodes, $\zeta$, gives this probability as $p = 1/(1 + e^{-\zeta})$. This output channel is trained on every training outage, including the unrecoverable ones. A cross-entropy term of weight 0.1 is used, the standard loss for a yes-or-no prediction. However, the trajectory part of the loss only uses the recoverable outages with a unique label. In the late-arrival study, the joint operator also has the following inputs: the slip, a validity mask and the time as a fraction of the original transfer time. The time fraction runs past one after the original arrival. The late arrival adds nodes to the re-planned trajectory, so every case is padded to 329 nodes, which is the longest one. A mask marks the real nodes.

Each ensemble has $M = 10$ members \citep{lakshminarayanan2017}, trained with the settings in \cref{tab:hyper}. Their mean is the prediction while their spread, $\sigma_{\mathrm{ens}}$, gives the uncertainty,
\begin{equation}
  \bar{\mathbf{u}} = \frac{1}{M}\sum_{j=1}^{M}\hat{\mathbf{u}}_j, \qquad
  \sigma_{\mathrm{ens}} = \frac{\lVert \operatorname{std}_j \hat{\mathbf{u}}_j \rVert}{\lVert \bar{\mathbf{u}} \rVert}.
  \label{eq:ensemble}
\end{equation}
As the spread is divided by the size of the prediction, it is on the same scale as the relative error of \cref{eq:loss}. The spread tends to be large where the answers are least reliable (\cref{sec:fixed}).

\subsection{Protocol}\label{sec:protocol}

Each study follows a plan that is fixed in advance of computing labels. This plan decides which outages count, how the error is measured, which rivals the operator must beat, and which transfers are held out for testing. For each method, the error on the predicted outage re-plan is the relative $L_2$ error of its controls against the unique exact re-plan. In both studies, the methods are compared using the severe outages. These are defined to be recoverable outages that remove the nominal trajectory's planned $\Delta v$ over the training outages by at least 51~m/s at a fixed arrival, and 61~m/s at a late one, which are medians over the training outages. The less severe outages are left out, as a trial run before the study showed that the cheap rivals come within 3\% of the exact re-plan after the mild outages. We compare the operator's median error with that of the best rival, defined as the rival with the lowest median error on the severe test outages. We then report the ratio of the two medians and the win rate, which we define as the share of severe outages on which the operator has a lower error. The results come from 30 test transfers with 750 outages in total for the fixed arrival and 1{,}498 for the late arrival, which neither the operator nor the learned correctors had seen before.

A different choice of 30 test transfers would give different values. To capture this variability, the main metrics are given with a 95\% interval. The interval is estimated via the bootstrap method \citep{efron1979}. 30 transfers are drawn randomly from the 30 test transfers, with repeats allowed, so that some transfers appear multiple times and some none. The metrics are then recomputed from the outages of the drawn transfers, and this is repeated 2,000 times. The interval reported spans the middle 95\% of results. We draw whole transfers rather than single outages, as outages on the same transfer tend to have similar errors.

\section{Results}\label{sec:results}

\subsection{Fixed arrival}\label{sec:fixed}

\begin{figure}[t]
  \centering
  \includegraphics[width=\textwidth]{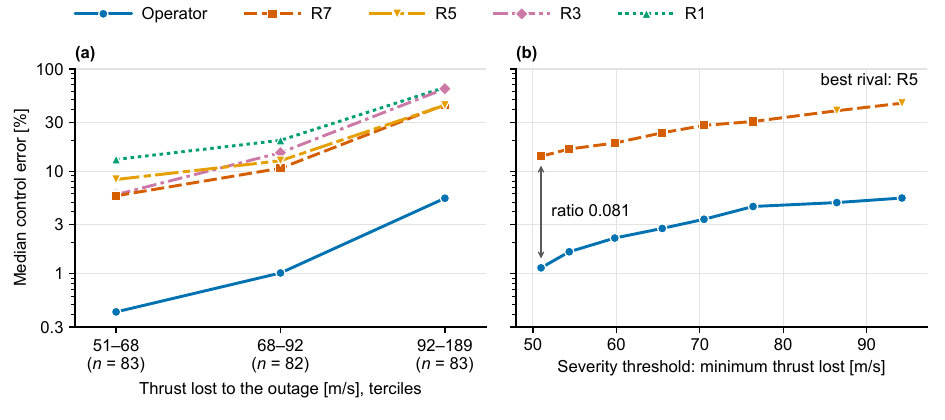}
  \caption{Median control error on the severe test outages at the fixed arrival.   (a) For the mixed operator and the rivals, with the outages split into three equal groups by the planned $\Delta v$ they remove.
  (b) The mixed operator and the best rival at each threshold of minimum thrust lost.}
  \label{fig:accuracy}
\end{figure}

The exact re-plans are repeatable, as 98\% of recoverable outages have a unique solution. On the 248 severe test outages, the mixed operator has a median control error of 1.14\%, against 14.0\% for R7, the best rival (\cref{tab:main}). The ratio of the two medians is 0.081 [0.068, 0.112] and the operator has the lower error on 99.6\% of these outages. Across all 524 recoverable test outages with a unique label, the errors are smaller, but the gap between the operator and the rivals remains, with 0.25\% against 3.9\% for R3, which is the best rival across this set. The advantage continues to hold as the outages increase in severity. The ratio stays within 0.073--0.123 across the severity terciles (\cref{fig:accuracy}a). On the outages that remove at least 94~m/s of planned $\Delta v$, the operator gives 5.5\% against 46\% for R5 (\cref{fig:accuracy}b).

Similarly, the operator is better at predicting the extra $\Delta v$. The quantity considered here is what would be used by a margin study, the re-plan's $\Delta v$ minus the nominal trajectory's. Its error per outage is 0.14~m/s at the median and 7.2~m/s at the 90th percentile, against 2.2 and 58~m/s for R3, on all 524 recoverable test outages with a unique label. Finally, the ensemble spread is a useful metric, as it indicates the predictions with greater errors. The Spearman rank correlation between the spread and the error is 0.91. When we solve exactly the 31\% of severe outages with the largest spread, the 90th-percentile error falls from 8.5\% to 2.3\%, suggesting that most of the large errors occur in cases where the operator members disagree.

When the R1/R3 solver inputs are removed, the operator maintains its accuracy on the severe outages, and a query becomes 16 times faster. On the severe test outages, the no-solver operator gives a median error of 1.11\%, against 1.14\% for the mixed operator. The two have no significant difference within the uncertainty, with a ratio of 0.98 [0.71, 1.24]. The joint operator, which also predicts whether a recovery exists, gives 1.20\% (\cref{tab:main}).

In contrast, the no-solver operator's error is 0.49\% on all recoverable outages, double the mixed operator's 0.25\%. This is likely due to R1/R3 being already accurate for mild outages, meaning that the operator has to learn only a small correction. The most significant price of the increase in speed for the no-solver operator is a weaker uncertainty estimate. Its spread ranks the errors with a Spearman rank correlation of only 0.66. Moreover, the accuracy of both operators improves with the number of labels. The mixed operator's error falls from 2.71\% with 5 labels per training transfer to 1.14\% with 17.5, from 25 drawn outages (\cref{tab:scaling}). Therefore, more labels would likely reduce the error further. The late-arrival study uses the joint operator, as it needs no solver at any slip, R3 cannot be computed with slip, and it also predicts whether a recovery exists.

\begin{table}[t]
\centering
\caption{Fixed-arrival re-planning results on the 30 unseen test transfers. Error is the relative $L_2$ error of the controls in percent. The ratio is the operator's median error over the rival's. The win rate is the share of severe outages where the mixed operator has the lower error. Query is the time for one query on one thread, which also includes solver steps.}
\label{tab:main}
\small
\begin{tabular}{@{}lrrrlrr@{}}
\toprule
 & \multicolumn{2}{c}{Severe outages} & All & & & \\
\cmidrule(lr){2-3}\cmidrule(lr){4-4}
Method & median & p90 & median & Ratio [95\% CI] & Win rate & Query \\
\midrule
Mixed operator & \textbf{1.14} & 8.47 & 0.25 & -- & -- & 355~ms \\
R0 & 61.75 & 98.38 & 38.28 & 0.018 [0.013, 0.028] & 1.000 & $<$1~ms \\
R1 & 20.69 & 85.51 & 7.54 & 0.055 [0.042, 0.079] & 0.992 & 55~ms \\
R2 & 44.75 & 105.57 & 6.38 & 0.025 [0.019, 0.039] & 0.915 & 123~ms \\
R3 & 16.84 & 95.67 & 3.91 & 0.067 [0.053, 0.094] & 0.984 & 277~ms \\
R5 & 15.15 & 56.24 & 5.88 & 0.075 [0.060, 0.099] & 1.000 & 55~ms \\
R6 & 26.43 & 71.22 & 8.88 & 0.043 [0.034, 0.060] & 1.000 & 123~ms \\
R7 & 14.02 & 69.20 & 3.98 & 0.081 [0.068, 0.112] & 0.996 & 277~ms \\
\midrule
No-solver operator & 1.11 & 6.90 & 0.49 & -- & -- & 23~ms \\
Joint operator & 1.20 & 6.59 & 0.55 & -- & -- & 22~ms \\
\bottomrule
\end{tabular}
\end{table}

\subsection{Recoverability}\label{sec:recov}

217 of the 750 test outages cannot be recovered on time, corresponding to 17\% of outages at flight-like frequencies. The joint operator can distinguish the unrecoverable outages from recoverable ones with an AUC of 0.999 (\cref{tab:recoverability}). This is the probability that a recoverable outage scores higher than an unrecoverable outage. To give a yes-or-no answer, a threshold on the recovery probability $p$ must be set. Above this threshold, the outage is deemed to be recoverable. It is set so that at most 5\% of the unrecoverable validation outages are called recoverable, which is the unsafe error. On the test outages, the joint operator has a false-recoverable rate of 4.1\% [1.4, 7.1] and a false-unrecoverable rate of 0.6\%. A tree model, which is extra trees fitted to the outage description of the training outages, performs worse, achieving 5.5\% and 2.3\%, respectively. Simpler predictors perform even worse, as a logistic rule on the time left and the outage length calls 9.9\% of the recoverable outages unrecoverable. Similarly, R3's failure flag achieves 9.7\% and 12.0\%, respectively.

\begin{figure}[t]
  \centering
  \includegraphics[width=\textwidth]{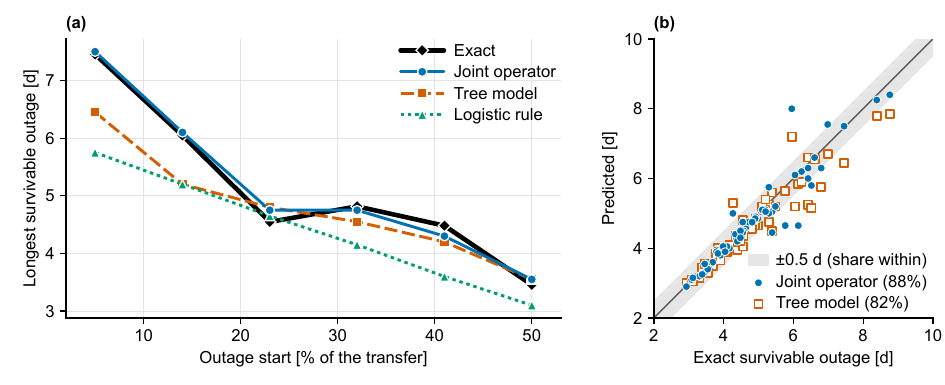}
  \caption{Longest survivable outage at the fixed arrival. (a) Against the outage start, on one test transfer. (b) Against the exact value, on all 10 transfers.}
  \label{fig:survival}
\end{figure}

The recoverability output can also be used to estimate the longest survivable outage at a given start time (\cref{fig:survival}). For a fixed start time, the operator is queried with outage lengths spaced 0.05~d apart, until the point where the recovery probability drops below 0.5. We compare the output of the operator with exact halving searches at six start times on 10 of the 30 test transfers. The exact longest survivable outage ranges from 2.9 to 8.8~d. The median error of the operator against the exact value is 0.10~d against 0.18~d for the tree model (\cref{tab:recoverability}). Being optimistic in the estimate is the unsafe direction. The operator is optimistic by more than half a day at only 3 of the 60 points, all starting in the first 14\% of the transfer. Calculating the exact point takes 8 solves and about 35~min at the median, against 1.4~s for the operator.

\begin{table}[t]
\centering
\caption{Recoverability predictions at the fixed arrival. Left: the 750 test outages, with 217 of them unrecoverable. FR is the false-recoverable rate and FU the false-unrecoverable rate, both at the 5\% validation threshold. Right: the longest survivable outage at 6 start times on 10 of the 30 test transfers, against exact searches. Optimistic is the share of points where it is more than half a day too long.}
\label{tab:recoverability}
\small
\begin{tabular}{@{}llrrrrr@{}}
\toprule
 & \multicolumn{3}{c}{Recoverable?} & \multicolumn{3}{c}{Survivable outage [d]} \\
\cmidrule(lr){2-4}\cmidrule(lr){5-7}
Predictor & AUC [95\% CI] & FR [\%] & FU [\%] & median err. & p90 err. & optimistic [\%] \\
\midrule
Joint operator & 0.999 [0.998, 1.000] & 4.1 & 0.6 & 0.10 & 0.58 & 5 \\
Tree model & 0.993 [0.986, 0.998] & 5.5 & 2.3 & 0.18 & 0.92 & 3 \\
Logistic rule & 0.975 [0.964, 0.986] & 8.3 & 9.9 & 0.35 & 1.18 & 8 \\
R3 failure flag & 0.892 [0.867, 0.915] & 9.7 & 12.0 & -- & -- & -- \\
\bottomrule
\end{tabular}
\end{table}

\subsection{Late arrival}\label{sec:late}
The late-arrival study investigates whether one operator, with the slip as an input, can re-plan an outage to any arrival up to 10~d late. The joint operator is compared with R0--R2, R5 and R6, as R3 cannot be computed with slip (\cref{sec:rivals}). 98.7\% of the 1{,}192 recoverable test outages have a unique exact re-plan, close to the 98.3\% in the fixed-arrival study. For the 596 severe test outages, the joint operator gives a median error of 1.67\% against 23.2\% for the best rival, R5. This corresponds to a ratio of 0.072 [0.062, 0.090], and the operator has the lower error on 98.7\% of them (\cref{tab:slip}). We also observe that the error stays flat across the slip, at 1.4--1.8\% in each slip range. In contrast, the error of R1 rises from 30.6\% to 38.2\%. To compare this operator with the mixed operator from the fixed-arrival study, we consider the outages starting within 5--50\% of the transfer. Here the ratio of the late-arrival operator is 0.065, against 0.081 at the fixed arrival, although the rivals differ. Every method has the largest errors with the earlier outages. The operator gives 2.19\% for starts at 5--25\% against 1.17\% at 50--75\%. This is likely because an early outage leaves more of the transfer to re-plan. Finally, five-member ensembles trained on 12 and 25 outages per transfer give 3.3 and 1.8 times the error of the full ensemble trained on 50 (\cref{tab:slipsecondary}). This suggests more labels would likely still improve the operator.

At a late arrival, recoverability depends on the slip, as an outage that cannot be recovered on time may still be recovered by arriving later. Of the 1{,}498 test outages, 306 are unrecoverable at their drawn slip, and the joint operator separates them with an AUC of 0.997. At the threshold that keeps the validation false-recoverable rate at 5\%, it has a false-recoverable rate of 3.6\% and a false-unrecoverable rate of 0.8\% on the test outages, against 5.6\% and 6.2\% for the tree model. However, at a 0.5 cut, the false-recoverable rate is about 1\% up to 5~d of slip and rises to 9.9\% above, where 9 of the 91 unrecoverable outages are called recoverable. Every predictor does worse at these large slips, with 28.6\% for the tree model and 44.0\% for the logistic rule. This may be because the few outages that stay unrecoverable after so much slip lie close to the boundary between the recoverable and unrecoverable outages. Consequently, a recoverable answer above 5~d of slip should be confirmed by an exact solve.

\begin{figure}[t]
  \centering
  \includegraphics[width=\textwidth]{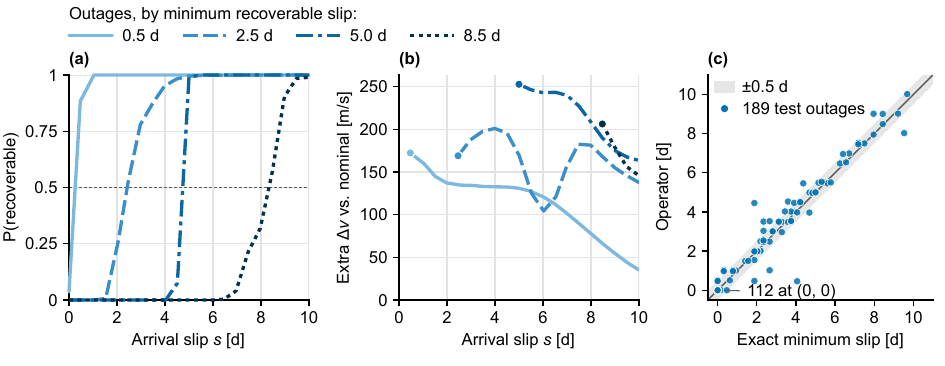}
        \caption{Late arrival, for the joint operator. (a) Recovery probability and (b) extra $\Delta v$ against the slip for four test outages. (c) Minimum recoverable slip against exact searches.}
  \label{fig:slip}
\end{figure}

We query the operator at 21 slips between 0 and 10~d to obtain the lateness curve of the outage. This curve is defined to be the change in recovery probability and the extra $\Delta v$ against the slip. It is obtained in 1.1~s with the operator, compared to 0.2~CPU-h for the exact solves (\cref{fig:slip}a, b). The minimum recoverable slip is defined to be the smallest slip on this 0.5~d grid at which the recovery probability reaches 0.5, so the grid can only make it later, which is the safe side. Of the 200 test outages queried this way, the operator finds 95\% recoverable within 10~d. Again, we compare it with exact halving searches on a second set of 200 test outages, 180 of them spread evenly over the start times, which take 872 solves in total (\cref{fig:slip}c). The operator is optimistic by more than half a day on 6 of them, all among the 84 that need a slip. These cases are mostly long outages and early starts.

For 189 outages, both the operator and the exact search find a minimum slip of 10~d or less. In these cases, the absolute error is 0.00~d at the median and 0.51~d at the 90th percentile. The larger errors tend to be the operator estimating a greater slip than the exact solution, the safe error. Finally, for 12 test outages solved exactly at every whole-day slip from 0 to 10~d, the operator correctly estimates the recoverability at 97\% of the points. For the recoverable outages, the extra $\Delta v$ error is 1.8~m/s at the median and 22~m/s at the 90th percentile.

\begin{table}[t]
\centering
\caption{Joint operator performance at a late arrival on the 30 unseen test transfers. Median control error [\%] on the 596 severe outages, and on all 1{,}177 recoverable outages with a unique label.}
\label{tab:slip}
\small
\begin{tabular}{@{}lrrrrr@{}}
\toprule
 & \multicolumn{4}{c}{Severe outages} & All \\
\cmidrule(lr){2-5}\cmidrule(lr){6-6}
 & & \multicolumn{3}{c}{slip [d]} & \\
\cmidrule(lr){3-5}
Method & median & 0--2 & 2--5 & 5--10 & median \\
\midrule
Joint operator & \textbf{1.67} & 1.73 & 1.40 & 1.77 & \textbf{0.96} \\
R0 & 87.79 & 70.38 & 82.86 & 90.47 & 85.42 \\
R1 & 36.31 & 30.56 & 32.49 & 38.18 & 21.28 \\
R2 & 79.23 & 60.57 & 68.71 & 85.26 & 79.08 \\
R5 & 23.17 & 23.41 & 19.22 & 24.58 & 13.15 \\
R6 & 45.41 & 36.70 & 42.71 & 48.47 & 39.40 \\
\bottomrule
\end{tabular}
\end{table}

\section{Discussion and limitations}\label{sec:discussion}

The operator only pays off in large outage studies. Before the first query, it costs 146.5~CPU-h for computing the training labels and 24.6~CPU-h for the training. On the other hand, solving all 750 test outages exactly takes only 14.8~CPU-h. To compare the operator to the exact-solve method, we run a design study as a mission designer might do. This is a margin study on the 750 test outages weighted to flight-like frequencies (\cref{tab:design}). In the operator route, the joint operator answers every outage, and the 10\% of outages it is least sure of are solved exactly. This results in the same probability of recovering on time as the exact route, and its 50th, 90th and 99th percentiles of extra $\Delta v$ are wrong by at most 1.6~m/s. It breaks even in total CPU-h required with the exact-solve method at about 11{,}400 queries. Below this number, the study is cheaper with the exact-solve route. However, studies over one family share the same labels, and the operator alone can answer $10^5$ more outages in under an hour.

A potential weakness of the operator is that it does not return a converged trajectory, and its guidance re-plan misses the DRO by a median of 1{,}360~km when flown open loop. This is about 8~m/s to correct, but the best rival's median miss is 20{,}900~km. We do not view this as a major problem, as a re-plan would be flown in closed loop in practice. Additionally, 8~m/s is small compared to the flight-weighted 99th-percentile margin of 210~m/s. Still, the operator cannot replace an exact solver when a converged trajectory is required. A possible application is an IPOPT warm start. When the joint operator is used with the nominal trajectory multipliers, IPOPT reaches an exact re-plan with a median solve time about 17\% less and cuts the 90th percentile from 24.2 to 8.2~s.

The results hold for one transfer family, with a specific dynamical model and objective. The re-plans minimise control energy, as the 32 lowest Fourier modes can represent the resulting smooth thrust histories sufficiently well. In contrast, a propellant-optimal re-plan switches the thrust fully on and off. Moreover, the dynamics are the BCR4BP with a constant thrust bound and no change in mass. In reality, none of these is true. The outage statistics may also not represent cislunar missions, as they are pooled from deep-space missions \citep{hooke2020}.

\section{Conclusion}\label{sec:conclusion}
We have shown that one neural operator, trained on exact re-plans, can accurately predict the outage recoverability and the re-planned trajectory for a family of NRHO-to-DRO transfers without a solver input. At the original arrival, the operator's median control error on the severe test outages is about a twelfth of the best cheap re-planner's, and it predicts whether a recovery exists with an AUC of 0.999. The same output can be used to predict the longest survivable outage at a given start to within 0.10~d at the median in about a second, where the solver route would take half an hour per point. With the slip as an input to the operator, the error is about a fourteenth of the best rival's at any slip up to 10~d, and the minimum recoverable slip is optimistic by more than half a day on only 6 of the 84 checked outages needing slip. The operator suits large and repeated margin studies over one transfer family. When the 10\% of outages the operator is least sure of are solved exactly, it reproduces the exact margins within 1.6~m/s and becomes cheaper than solving every outage after about 11{,}400 queries. Future work could include passing the operator's output to an exact solver that performs a propellant-optimal re-plan, and re-planning in closed loop after every outage instead of once. Training operators on other transfer families would also show how the method generalises beyond the NRHO-to-DRO family.

\section*{Acknowledgements}
Generative AI tools were used for coding assistance and to reformat the LaTeX source file. The author designed the study and checked all outputs. The author takes full responsibility for the content.

\bibliographystyle{unsrtnat}
\bibliography{references}

\appendix
\section{Problem details}\label{app:problem}
For the whole study, the dynamics model used is the BCR4BP in the Earth--Moon rotating frame, with lengths in LU (384{,}400~km), times in TU (375{,}190~s) and the mass parameter $\mu = 0.0121506$. The Sun moves on a circle of radius $a_s = 388.81$~LU about the Earth--Moon barycentre. The Sun's phase is $\theta_s = \theta_0 + \omega_s t$, where $\omega_s = -0.92520$, and its mass is $m_s = 328{,}900.54$ in Earth--Moon units. The model follows the BCR4BP formulation in \citep{boudad2020}, and with the length unit in this paper, the Sun's distance is 0.1\% below one astronomical unit. Given the thrust acceleration $\mathbf{u}$, the equations of motion are
\begin{equation}
  \ddot{x} - 2\dot{y} = \frac{\partial \Omega}{\partial x} + u_x, \qquad
  \ddot{y} + 2\dot{x} = \frac{\partial \Omega}{\partial y} + u_y, \qquad
  \ddot{z} = \frac{\partial \Omega}{\partial z} + u_z,
  \label{eq:bcr4bp}
\end{equation}
\begin{equation*}
  \Omega = \frac{x^2 + y^2}{2} + \frac{1 - \mu}{r_1} + \frac{\mu}{r_2} + \frac{m_s}{r_s}
  - \frac{m_s}{a_s^2}\left(x \cos\theta_s + y \sin\theta_s\right),
\end{equation*}
where $r_1$, $r_2$ and $r_s$ are the distances to the Earth, the Moon and the Sun. $\lVert\mathbf{u}\rVert$ is bounded by 0.82~mm/s$^2$ times the available thrust. The nominal trajectory is designed with the thrust limited to 90\% of this bound. On the other hand, the re-plan is allowed to use the full bound, so this reserve thrust can be used to make up the thrust lost during an outage.

We use symmetric differential correction to compute both the departure NRHO and the arrival DRO in the CR3BP. The NRHO has $z = -0.1821$~LU where it crosses the $xz$-plane at apolune, and the DRO crosses the Earth--Moon line 0.17~LU from the Moon on the Earth side. We sample 270 transfers between these orbits with a scrambled Sobol sequence over the transfer time, the departure and arrival phases and the Sun phase. The test transfers are kept inside the training range, as the first three parameters are sampled over the test envelope widened by 5\% of each range on either side. This changes their ranges from 15.6--19.1~d, 0.42--0.58 of the NRHO period from apolune and $\pm0.08$ of the DRO period to 15.425--19.275~d, 0.412--0.588 and $\pm0.088$. The transfers outside the envelope are always used for training. The Sun phase covers one period of the solar tide, 0--180$^\circ$, so it is not widened. Each transfer is transcribed by multiple shooting on 200 nodes, and the dynamics are integrated with RK4 at a step of 0.003~TU. We scale this step by $(r_2/40{,}000~\mathrm{km})^{1.5}$ within 40{,}000~km of the Moon, as the faster motion near the Moon requires finer steps. The worst terminal position error is 0.023~km against an integration with a four times finer step. We also check the 100~km clearance above the lunar surface.

We build the thrust history by drawing one outage, and then adding a second and a third outage independently with probabilities of 0.15 and 0.03. The outages start at a uniformly drawn time within 5--50\% of the transfer in the fixed-arrival study and 5--75\% in the late-arrival study. Their durations follow a Weibull distribution with a shape of 1.17 and a scale of 3.6~d, and each duration is redrawn until it lies within 0.5--10~d. The scale is 1.5 times the flight value of 2.41~d. Future outages are unknown to the operator. In 30\% of histories, the available thrust falls linearly from full at departure to a loss drawn uniformly in 0--20\% at the original arrival and continues at the same rate after it. For each case, we also store a weight, which is how likely the case is under the flight reference divided by how likely it is in training. This enables us to re-weight the results to flight-like frequencies. The flight reference uses the flight scale of 2.41~d and probabilities of 0.05 and 0.002 for the second and third outages. Consequently, only the number and length of the outages change, and the start window and thrust decline do not. With these weights, the 524 recoverable test outages of the fixed-arrival study with a unique label give estimates about as precise as 448 equally weighted outages.

\section{Rivals and operator}\label{app:rivals}

R0 does not re-plan, and flies the nominal controls clipped radially to the available thrust. R1 solves one convex sub-problem with the dynamics linearised about the nominal. Its trust region is made large enough for the result to be the exact optimum of the linearised re-plan. R2 takes two SCvx iterations from the nominal \citep{malyuta2022}. A step is accepted only if a non-linear merit function falls by at least a tenth of the predicted amount. The trust region starts at a radius of 0.05, and doubles after good steps and halves after rejected ones. Both R1 and R2 use virtual controls and a slack on the terminal state, ensuring every sub-problem has a solution. R3 takes one SQP step from the nominal and its multipliers, while using the exact Hessian of the Lagrangian. The step is solved as a quadratic program with IPOPT. As the thrust bound is linearised in this step, the results can go above the bound. Therefore, the thrust is projected radially back onto the bound. For the late-arrival study, we extend the nominal trajectory by coasting along the DRO for the extra number of days. After the re-plan point, R0 flies this extended nominal trajectory without clipping it to the declining thrust. However, the extended nominal is not a solution of the longer problem, meaning that it has no multipliers. This rules out R3 and R7 in the late-arrival study.

R5--R7 add a learned correction to R1--R3. These corrections are the differences between the label and the rival's controls at each node. We compress the correction histories of the training and validation outages by principal component analysis (PCA) \citep{jolliffe2016} to 20 components, reducing the number of outputs. Then, 400 extra trees \citep{geurts2006} predict these components from the outage description, both using the implementations of scikit-learn \citep{pedregosa2011}. For recoverability, we compare the operator with a tree model and a logistic rule, both fitted on the training outages. The tree model uses 500 extra trees on the outage description and on features of the first outage. On the other hand, the logistic rule, a logistic regression \citep{cox1958}, uses only the time left after the first outage and its length. The slip is added to this for the late-arrival study. The outage description used by R5--R7 and the tree model includes every drawn outage window, despite the later ones being unknown at re-plan time. However, this does not give these rivals useful information, as the labels do not depend on the later outages. We train each operator member with the settings of \cref{tab:hyper}.

\begin{table}[h]
\centering
\caption{Operator and training settings of all three operators.}
\label{tab:hyper}
\small
\begin{tabular}{@{}l>{\raggedright\arraybackslash}p{0.72\textwidth}@{}}
\toprule
Architecture & Fourier neural operator, width 64, 32 modes, 4 layers \\
Ensemble & 10 members: seeds 0--4 of each kind (mixed), seeds 0--9 (no-solver and joint) \\
Training & 1{,}500 epochs, batch 32, Adam, learning rate $2\times10^{-3}$, weight decay $10^{-5}$, cosine schedule \\
Selection & epoch with the lowest validation median error, checked every 25 epochs \\
Recoverability output & joint operator only; one extra output channel, whose mean over the valid nodes is the logit;
cross-entropy weight 0.1 \\
\bottomrule
\end{tabular}
\end{table}

\begin{table}[h]
\centering
\caption{Fixed arrival. Median error [\%] on the severe test outages with less training data or fewer epochs.}
\label{tab:scaling}
\small
\begin{tabular}{@{}llrr@{}}
\toprule
Setting & Labels / transfers & Mixed operator & No-solver operator \\
\midrule
5 labels / transfer & 1{,}050 / 210 & 2.71 & 3.72 \\
10 labels / transfer & 2{,}098 / 210 & 1.86 & 1.98 \\
15 labels / transfer & 3{,}056 / 210 & 1.38 & 1.28 \\
25\% of transfers & 920 / 52 & 3.37 & 4.79 \\
50\% of transfers & 1{,}843 / 105 & 1.96 & 2.34 \\
500 epochs & 3{,}672 / 210 & 1.34 & 1.27 \\
Full (1,500 epochs) & 3{,}672 / 210 & 1.14 & 1.11 \\
\bottomrule
\end{tabular}
\end{table}
\begin{table}[h]
\centering
\caption{Late arrival. (a) Median error [\%] on the severe test outages with less training data. (b) FR and FU rates at stricter validation targets.}
\label{tab:slipsecondary}
\small
\begin{subtable}[t]{0.58\textwidth}
\centering
\caption{}
\begin{tabular}{@{}rrrr@{}}
\toprule
Outages / transfer & Members & Median [\%] & Ratio to full \\
\midrule
50 & 10 & 1.67 & 1 \\
25 & 5 & 3.07 & 1.838 \\
12 & 5 & 5.52 & 3.306 \\
\bottomrule
\end{tabular}
\end{subtable}%
\hfill
\begin{subtable}[t]{0.32\textwidth}
\centering
\caption{}
\begin{tabular}{@{}lrr@{}}
\toprule
Target & FR [\%] & FU [\%] \\
\midrule
2\% & 1.3 & 2.9 \\
1\% & 1.0 & 3.9 \\
\bottomrule
\end{tabular}
\end{subtable}
\end{table}

\section{Cost and timing}\label{app:cost}
In \cref{tab:design}, the cost of the operator includes its labels and training while counting every exact solve. \cref{tab:timing} gives the query times behind it, measured on one thread of an AMD EPYC 9R14 processor. The exact and rival times are those measured when the labels were solved.

\begin{table}[h]
\centering
\caption{Design study on the 750 test outages, weighted to flight-like frequencies. The joint operator answers every outage, and the share it is least confident of is solved exactly. Errors are relative to the exact solves.}
\label{tab:design}
\small
\begin{tabular}{@{}lrrrr@{}}
\toprule
Route & Exact solves & CPU-h & P(recover) error [pts] & 99th pct.\ error [m/s] \\
\midrule
Exact only & 750 & 14.8 & 0 & 0 \\
Operator, 0\% checked & 0 & 0.0 & +0.6 & +3.8 \\
Operator, 2\% checked & 15 & 0.8 & +0.8 & +3.8 \\
Operator, 5\% checked & 38 & 1.9 & +0.6 & +7.3 \\
Operator, 10\% checked & 75 & 3.5 & +0.0 & +0.2 \\
Operator, 20\% checked & 150 & 5.9 & +0.0 & +0.2 \\
Operator, 50\% checked & 375 & 12.2 & +0.0 & +0.2 \\
\bottomrule
\end{tabular}
\end{table}
\begin{table}[h]
\centering
\caption{Query times [ms] on one thread. The mixed operator's single query includes solving R1 and R3. The batch column counts the neural operator network only. A single exact IPOPT solve from the nominal takes 3.3~s.}
\label{tab:timing}
\small
\begin{tabular}{@{}lrr@{}}
\toprule
Operator & One query & Per query in a batch \\
\midrule
Mixed operator & 355.0 & 8.57 \\
No-solver operator & 22.5 & 10.37 \\
Joint operator & 22.5 & 7.32 \\
\bottomrule
\end{tabular}
\end{table}

\end{document}